\documentstyle[12pt]{article}
 
\begin{document}
\vspace{-2.0cm}
\bigskip
 
\begin{center}
{\Large \bf Collective coordinate quantization of $CP^1$ model coupled
to Hopf term revisited}
\vskip .8 true cm
{\bf Biswajit Chakraborty}\footnote{biswajit@bose.res.in} 
\vskip 1.0 true cm

S. N. Bose National Centre for Basic Sciences \\
JD Block, Sector III, Salt Lake City, Calcutta -700 098, India.
\end{center}
\bigskip
 
\centerline{\large \bf Abstract}
We show that the system where $CP^1$ model coupled to Hopf term can
reveal fractional spin in a collective coordinate quantization scheme,
provided one makes a transition to physically inequivalent sector
within a same solitonic sector characterized by a nonvanishing
topological number 

Ever since Wilczek and Zee [1] showed that the system involving $O(3)$
non-linear sigma model (NLSM) coupled to the Hopf term can exhibit fractional
spin in a quantum mechanical analysis using path integral technique, numerous
attempts have been made to obtain the same result in a canonical Hamiltonian
framework. One of the first attempts in this direction was made by Bowick,
Karabali and Wijewardhana (BKW)[2]. However their analysis was incomplete
because of the following reasons. Firstly the model was {\it altered}
using certain identity, which does not correspond to a constraint of the
theory. Besides, this identity was valid only in the radiation gauge. 
Secondly, the entire analysis was carried out at the
classical level; the Dirac brackets (DB) were not elevated to quantum
commutators[3]. In fact the structures of the Dirac brackets are too complicated
(as they involve product of field variables and beset with operator ordering
ambiguities) to lend themselves to be elevated to quantum commutators.
This was shown in [4] in an equivalent formulation using $CP^1$ variables.
This is very important as, strictly speaking, fractional spin has a
quantum origin and stems from the multiply-connected nature of the
configuration space[5]. Of course the determination of the spectrum of angular 
momentum operator is a quite nontrivial in a completely field theoretic
set-up for which they take recourse to
collective coordinate quantization, where the problem is essentially reduced
to a quantum mechanical problem of a rigid rotor. But again here they consider
only the, above mentioned, {\it altered} model.

One can nevertheless work with a definition of fractional spin $(J_f)$, which
can be defined at the classical level itself. It was also, to the best
of our knowledge, was first introduced by BKW themselves in [2]. They
essentially compute the difference between the expression of angular
momentum $(J^s)$ obtained from the symmetric definition of energy-momentum
tensor (which is obtained by functionally differentiating the action
with respect to the metric and setting it flat eventually)
and the one obtained from Noether's prescription $(J^N)$,
$$ J_f= (J^s - J^N). \eqno(1)$$
The latter boils down to just orbital angular momentum for theories involving scalar
fields only, like NLSM fields considered in [2]. $J_f$ can therefore be
interpreted as fractional angular momentum of the $\it altered$ model
with certain justification. For theories involving
vector fields like Chern-Simons (CS) term etc, $J^N$ contains some additional
terms, apart from the orbital angular momentum, although a topological term
like CS/Hopf term do not contribute to the, above mentioned energy-momentum tensor.
In this case, $J_f$ turns out to be a 
nontrivial boundary term, if one carries out the entire analysis in a gauge
independent manner. Interestingly, one can show that this $J_f$ has only
a restricted gauge invariance like the CS action itself, in the sense that
both are gauge invariant under only those gauge transformations which
tend to identity asymptotically[6]. Evaluating $J_f$ in two distinct gauges
having different asymptotic behaviour, expectedly, will yield different results.
But this is hardly surprising, as the CS action itself has this restricted
symmetry and therefore these two gauges refer two inequivalent physical sectors.
Computing $J_f$ in radiation gauge yields the same result as can be obtained
through other approaches using radiation gauge right from the beginning[7,8].
This further justifies the interpretation of $J_f$ as fractional angular 
momentum and also the gauge independent expression for $J_f$ provides a kind
of master expression, from where we can access distinct physical sectors by
specifying the appropriate gauge fixing condition and compute the associated
fractional spin[9].

Formally, the Hopf action has the same mathematical form as that of CS action
and therefore enjoys the same kind of restricted gauge invariance.
The only difference being that the gauge fields appearing in Hopf term should
not be treated as independent degree of freedom, unlike that in CS action,
and should rather be eliminated in favour of the current by using certain
gauge fixing conditions. Typically, the Hopf term therefore represents a
gauge fixed non-local current-current interaction. That's what happens
precisely for O(3) NLSM. However, the equivalent $ CP^1$ model has a $U(1)$
symmetry and allows rewriting the Hopf term, using $CP^1$ variables, in
a {\it local} gauge invariant
manner-albeit in the above mentioned restricted sense. One can therefore
carry out a gauge independent analysis of the same and finds, surprisingly,
that fractional spin vanishes as $J^s$ and $J^N$ become identical [3,4], unless
the model is altered a la BKW [2] to reproduce their result of fractional spin.
This is of course a classical result and therefore does not rule out the
emergence of any fractional spin at the quantum level. Note that this has
to be of order $\hbar$, so that it can disappear at the classical limit
and necessarily be therefore different from BKW result as the Hopf parameter
itself has the dimension of $\hbar$. Again using Batalin-Tyutin scheme
of quantization [10], (where the second class constraints are elevated to the
level of first class constraints, so that operator ordering problems
can be avoided, as one just needs to elevate the basic Poisson brackets
to quantum commutators in order to carry out Dirac scheme of quantization
of the system)
it has been shown[11] that the angular momentum $J$ (Note that now $J=J^s=J^N$)
does not get any quantum correction unlike energy, where Hopf term 
, although a topological term, is
shown to induce finite energy density in the nontrivial topological sector
at the quantum level. Not only that, taking recourse to collective
coordinate quantization as was done in [2], and taking a profile for
the $CP^1$ fields, appropriate for the topological sector $Q=1$, it was
found to yield only integer spectrum for the angular momentum thus signaling
the absence of any fractional spin. On the other hand, it was observed in [12]
very recently that collective coordinate quantization can yield fractional
spin. The purpose of the letter ,in fact, was to resolve this {\it apparent}
contradiction
and put it in the perspective of other works carried out in this direction.

To begin with, let us consider the action $S$ of the theory, where $CP^1$
model $S_0$ has been coupled to the Hopf term $S_H$,
$$S= S_0+ S_H \eqno(2a)$$
where,
$$S_0= \int d^3x [(D_{\mu}Z)^{\dagger} (D^{\mu}Z)-{\lambda}(Z^{\dagger}Z-1)] \eqno(2b)$$
and 
$$S_H=\theta \int d^3x \epsilon^{\mu\nu\lambda}
[Z^{\dagger}{\partial}_{\mu}Z{\partial}_{\nu}Z^{\dagger}{\partial}_{\lambda}Z + h.c.] \eqno(2c)$$ 
Here $Z= {\pmatrix{z_1 \cr z_2}}$ is a doublet of complex scalar fields subjected
to the constraint $Z^{\dagger}Z=1$, which is enforced by the Lagrange multiplier $\lambda$
in (2b). This is supposed to capture the integer (Hopf number), associated to the
fundamental group $\Pi_1({\cal C}) = \Pi_3(S^2) = {\cal Z}$ of the configuration
space $\cal C$.

Clearly, two profiles of $CP^1$ fields $Z(x)$ and $Z'(x)$, which are related by
a $U(1)$ transformation as,
$$ Z(x) \rightarrow Z'(x)= e^{-i\alpha (x)}Z(x) \eqno(3)$$
although belong to the same solitonic sector (this captures the integer associated
with $\Pi_0({\cal C}) = \Pi_2(S^2) = {\cal Z} $)
, characterized by the topological index
$$Q=-{i\over {2\pi}}\int d^2x \epsilon^{ij}{\partial}_iZ^{\dagger}{\partial}_jZ \eqno(4)$$
and correspond to the same NLSM fields $n_a=Z^{\dagger}{\sigma}_aZ$, (obtained by
the Hopf map) they do not necessarily belong to the same physical sector, unless
one demands that $\alpha (x)\rightarrow 0$ at spacetime infinity. This is because, under
the transformation (3), the Hopf action (2c) does not remain invariant and undergoes
the following transformation:
$$S_H[Z]\rightarrow S_H[Z']=S_H[Z]-2i{\theta}\int d^3x \epsilon^{\mu\nu\lambda}
{\partial}_{\mu}({\alpha}{\partial}_{\nu}Z^{\dagger}{\partial}_{\lambda}Z) \eqno(5)$$
And this $\theta$ dependent term will vanish if and only if the gauge parameter
$\alpha (x)$ has this desired asymptotic property. To put it more precisely, let
$G = \{e^{-i\alpha (x)}\}$ be the set of all gauge transformation and let
$H = \{e^{-i\alpha (x)}\}$ be a subgroup of $G$, subject to the restriction that
$\alpha (x)\rightarrow 0$ at spacetime infinity. Then the coset $G/H$ splits 
into distinct equivalence classes of gauge transformations. Any pair of gauge 
transformations are equivalent if and only if they are identical asymptotically.
 And space of distinct physical sectors are in one-to-one correspondence with 
the elements of $G/H$.

From the point of view of collective coordinate quantization, it will be useful
to restrict the form of $\alpha (x)$ further, by taking it to depend on the time
variable $t$ only. With this (5)  reduces to,
$$S_H[Z]\rightarrow S_H[Z']=S_H[Z]-2i{\theta}\int d^3x {\dot\alpha}\epsilon^{ij} 
{\partial}_iZ^{\dagger}{\partial}_jZ \eqno(6)$$
Correspondingly, the total Lagrangian changes as,
$$L\rightarrow L'=L + 4\pi \theta Q {\dot\alpha} \eqno(7)$$
where we have made use of the fact that $S_0$ (2b) is invariant under gauge
transformation (3) and the relation (4). We thus observe that the gauge transformation
$$Z\rightarrow Z'=e^{-i\alpha (t)}Z \eqno(8)$$
induces a nontrivial transformation in the collective coordinate Lagrangian,
which again stems from the {\it non-invariance} of the Hopf action (2c),
and takes one from one physical sector to another physical sector {\it within}
the same topological sector with non-vanishing topological charge $Q$ (4).
 
In particular, in $Q=1$ sector, one can consider the following two configurations
of the $CP^1$ fields,
$$Z= \pmatrix{cos({g(r)\over 2})\cr sin({g(r)\over 2})e^{i(\phi +\alpha (t))}}\eqno(9a)$$
and
$$Z'=\pmatrix{cos({g(r)\over 2})e^{-i\alpha (t)}\cr sin({g(r)\over 2})e^{i\phi}}\eqno(9b)$$
related in the manner of (8).
Here the function $g(r)$ satisfies $g(0)=0$ and $g(\infty)=\pi$. While the configuration
(9a) was used in [11], the configuration (9b) was used in [12]. And it is
quite clear that while the Hopf action vanishes for the configuration (9a) [11], it
does not vanish for that of (9b). Using (7) and the results in [11], we find that
the total Lagrangian corresponding to the configurations (9a) and (9b) become,

$$L={\pi\over 2}{\lambda}{\dot\alpha}^2 - N \eqno(10a)$$
and 
$$L'={\pi\over 2}{\lambda}{\dot\alpha}^2+4\pi\theta{\dot\alpha}-N \eqno(10b)$$
where,
$$\lambda=\int dr r sin^2g(r)\eqno(11a)$$
and
$$N={\pi\over 2}\int dr r[(g'(r))^2+{1\over r^2}sin^2g(r)]\eqno(11b)$$

Following [2,11], it is now straightforward to show that while the spectrum for the
angular momentum for (10a) is given by,
$$J=integer  \eqno(12a)$$
the spectrum for (10b) is given by
$$J=integer + 4{\pi}{\theta} \eqno(12b)$$
Clearly, fractional spin depends on the presence of the term involving $\theta {\dot \alpha}$.

Here we would like to contrast this with the collective coordinate Lagrangian obtained from the
BKW altered model, which is given as [11],
$${\tilde L}={\pi\over 2}{\lambda}{\dot\alpha}^2+\theta{\dot\alpha}-N \eqno(13)$$
Although both of them have the $\theta$-dependent term present, the
accompanying coefficients are different. Consequently, the spectrum is simply
given by 
$$J=integer + {\theta} \eqno(14)$$  

This clearly demonstrates that fractional spin in collective coordinate
quantization can be obtained only by making such a gauge transformation
that one lands up in an inequivalent physical sector within the same
topological sector. In a sense, this is therefore analogous to the model
involving $SU(2)$ Chern-simons term, where asymptotically inequivalent
gauge conditions yield different result [9].
It is not difficult to generalize this result to arbitrary
topological sector $Q$, where fractional spin is given by ($\sim {\theta}Q^2$). 
Since the NLSM fields $n_a$ are $U(1)$ invariant,
and the Hopf term is defined in a particular gauge, as discussed earlier, one does
not have such a scope of connecting inequivalent physical sectors, when
the model is written in terms of NLSM fields. This is the advantage of using $CP^1$
variables. However it has been shown, by Kimura et.al [13], recently that this
($\sim {\theta}Q^2$) can be obtained using adjoint orbit parametrization albeit
for a restricted class of configurations. It was, in fact, argued in [13] that
(4) can reproduce the Hopf index only for the configurations in the trivial, i.e. vanishing
topological sector. Using adjoint orbit parametrization, however, the Hopf term
can be written in such a manner that it truly represents Hopf number to any
configurations. It will be interesting to relate the observations made in [13]
with the connection of fractional spin with asymptotically nontrivial gauge
transformation using $CP^1$ variable introduced here.

\end{document}